\documentclass{article}
\begin{document}
\title{Non-geodesic motion in $f({\mathcal G})$ gravity with non-minimal coupling}
\author{Morteza Mohseni\thanks{E-mail address:m-mohseni@pnu.ac.ir}\\
\small Physics Department, Payame Noor University, 19395-4697 Tehran, Iran}
\maketitle
\begin{abstract}
The dynamics of test particles in $f(\mathcal G)$ modified Gauss-Bonnet gravity is investigated. It is shown that in $f({\mathcal G})$ gravity 
models with non-minimal coupling to matter, particles experience an extra force normal to their four-velocities and as a result move along non-
geodesic world-lines. The explicit form of the extra force depends on the function of the Gauss-Bonnet term included in coupling 
term. The effects of this force on the relative accelerations of particles are studied.
\vspace{5mm}
\end{abstract}
\section{Introduction}
The $f({\mathcal G})$ modified Gauss-Bonnet gravity (or $f({\mathcal G})$ gravity for short) is among the plethora of models proposed recently 
to explain certain physical observations such as the late time accelerated expansion of the universe \cite{nojiri1}. It is well known that 
adding the Gauss-Bonnet curvature ${\mathcal G}$ to the usual scalar curvature in Einstein-Hilbert action does not alter the equations of motion 
of general relativity due to the fact that it is a topological invariant in four dimensions, even though it may have some other non-trivial 
contributions \cite{liko,olea}. However this term can be incorporated into space-time dynamics in several ways, namely by considering higher 
dimensions, e.g as in \cite{ching} or via non-minimal coupling to some scalar field in arbitrary space-time dimensions, which is also called 
modified Gauss-Bonnet gravity, as in string gravity, see e.g \cite{nojiri2,calcagni}. In $f({\mathcal G})$ gravity an arbitrary function of the 
Gauss-Bonnet curvature is added to the Einstein-Hilbert action resulting in modified equations of motion. Physical and astrophysical 
implications of such a modification have been extensively studied, e.g. in 
\cite{nojiri3,brevik,li,cognola2,bamba,bazeia,felice,goheer,zhou,bom}. An extension of the model has been considered in 
\cite{cognola,felice2,ali,ghal}. A review and comparison with $f(R)$ gravity model may be found in \cite{odinoj}. 

In the present work we aim to investigate the dynamics of test particles in $f({\mathcal G})$ gravity models. This is motivated by the essential 
role of particle  dynamics in physics and astronomy. In particular, in the present context, the particle dynamics may be used to put constraints 
on the form of the Lagrangian as it has been done in \cite{zakharov} for the case of $R^n$ gravity. We show that within the usual $f({\mathcal 
G})$ gravity, i.e. $f({\mathcal G})$ gravity minimally coupled to matter, test particles move along the background geodesics, but if we consider 
non-minimal coupling with matter extra force arises due to the non-minimal coupling resulting in non-geodesic motions. As it has been shown in  
\cite{bertolami}, a similar effect can emerge in models in which an arbitrary function of the scalar curvature is non-minimally coupled to 
matter in the framework of f(R) gravity \cite{odin1,odin2}. It should be noted that a non-minimal coupling does not necessarily result in non-
geodesic motion. This point has been shown in \cite{soti} for a $R$-matter coupling with various choices of the matter field.   

In this work we consider an action which consists of the usual $f({\mathcal G})$ gravity action and an arbitrary function of ${\mathcal G}$ non-
minimally coupled to matter Lagrangian. This is a slight modification to the model proposed in \cite{tret1,tret2} which has attracted some 
attention, namely in \cite{sade} in which a modified Gauss-Bonnet model with non-minimal coupling has been used to investigate the cosmic 
acceleration and crossing of phantom divide.

In what follows we first give a brief review of the $f({\mathcal G})$ gravity and by deploying the energy-momentum tensor of a perfect fluid we 
show explicitly that within the framework of $f({\mathcal G})$ gravity test particles move along the background geodesics. In the next section 
we show that in a $f({\mathcal G})$ gravity with non-minimal coupling to matter an extra force will be exerted on particles resulting in non-
geodesic motion. Then we obtain the world-line deviations of particles as a generalization of the geodesic deviation equation for both 
$f({\mathcal G})$ and $f(R)$ gravities with non-minimal coupling and compare them. We conclude the work with a discussion of the results.  
\section{Motion in $f({\mathcal G})$ gravity}
In $f({\mathcal G})$ gravity, the space-time equation of motion can be obtained from the action \cite{nojiri1}
\begin{equation}
S=\int{\sqrt -g}\left(\frac{1}{2\kappa^2}R+f({\mathcal G})+L\right)d^4x\label{h1}
\end{equation}
where $\kappa^2=8\pi G$ is the Einstein-Hilbert coupling constant which we set to unity henceforth, $L$ is the matter Lagrangian, and 
$f({\mathcal G})$ is an arbitrary function of ${\mathcal 
G}=R^2-4R_{\alpha\beta}R^{\alpha\beta}+R_{\alpha\beta\kappa\lambda}R^{\alpha\beta\kappa\lambda}$.
Varying this action with respect to metric components $g_{\mu\nu}$ leads to the equation of motion
\begin{equation}\label{h3}
T^{\mu\nu}=G^{\mu\nu}-g^{\mu\nu}f+4f^\prime H^{\mu\nu}
\end{equation}
where $G^{\mu\nu}$ is the Einstein tensor, $T^{\mu\nu}=\frac{2}{\sqrt -g}\frac{\delta(L{\sqrt -g})}{\delta 
g_{\mu\nu}}$ is the energy-momentum tensor, $f$ and $f^\prime$ stand for $f({\mathcal G})$ and $\frac{df({\mathcal G})}{d{\mathcal G}}$ 
respectively, and 
\begin{eqnarray}\label{e32}
H^{\mu\nu}&=&RR^{\mu\nu}+{R^\mu}_{\alpha\beta\gamma}R^{\nu\alpha\beta\gamma}-2R^\mu_\alpha 
R^{\alpha\nu}\nonumber\\&&+2R^{\mu\alpha\beta\nu}R_{\alpha\beta}-2G^{\mu\nu}\nabla^2-
R\nabla^\mu\nabla^\nu\nonumber\\&&-2g^{\mu\nu}R^{\alpha\beta}\nabla_\alpha\nabla_\beta+2R^{\alpha\nu}\nabla_\alpha\nabla
^\mu\nonumber\\&&+2R^{\mu\alpha}\nabla_\alpha\nabla^\nu-2R^{\mu\alpha\beta\nu}\nabla_\alpha\nabla_\beta.
\end{eqnarray}
Now we show explicitly that within the framework of $f({\mathcal G})$ gravity test particles move along the background geodesics. 
Taking the covariant derivative of both sides of equation (\ref{h3}) we can simplify the result to
\begin{equation}\label{h4}
\nabla_\nu T^{\mu\nu}=f^\prime\nabla^\mu{\mathcal G}-\nabla^\mu f
\end{equation}
where the right hand side vanishes identically. For a perfect fluid with pressure $p$ and energy density $\rho$ we have
\begin{equation}\label{h5}
T^{\mu\nu}=ph^{\mu\nu}+\rho u^\mu u^\nu
\end{equation}
where $h^{\mu\nu}=g^{\mu\nu}+u^\mu u^\nu$. Inserting this into (\ref{h4}) and projecting the resulting equation parallel and normal to $u^\mu$ 
results in the continuity equation  
\begin{equation}\label{h6}
{\dot \rho}=-(p+\rho)\theta
\end{equation}
where $\theta=\nabla_\nu u^\nu$, and the equation of motion
\begin{equation}\label{h7}
\nabla^\mu p+u^\mu{ \dot p}+(p+\rho){\dot u^\mu}=0
\end{equation}
where an over-dot means $u^\alpha\nabla_\alpha$ and the normalization $u_\mu u^\mu=-1$ has been used. For a perfect dust we have $p=0,\rho=nm$ 
where $n$ is the particles number density and $m$ is the mass of each particle and thus the last equation reduces to  
\begin{equation}\label{h8}
{\dot u^\mu}=0
\end{equation}
corresponding to a geodesic trajectory. For massless particles we have $u_\mu u^\mu=0$ and the above procedure again results in geodesic 
equation, as in general relativity.
\section{Non-geodesic motion}
In this section we consider an extension of the above model in which an arbitrary function of the Gauss-Bonnet curvature is non-minimally 
coupled to matter. The action may be written in the following form
\begin{equation}
S=\int\hspace{-1mm}{\sqrt -g}\left(\frac{1}{2}R+f({\mathcal G})+(1+\lambda F({\mathcal G}))L\right)d^4x\label{h14}
\end{equation}
in which $F({\mathcal G})$ is an arbitrary function of ${\mathcal G}$ and $\lambda$ is a coupling constant. 
This may be compared with the action
\begin{equation}
S=\int{\sqrt -g}\left(\frac{1}{2}R+f({\mathcal G})L\right)d^4x\label{h14a}
\end{equation}
introduced in \cite{tret1,tret2}. The equation of motion resulting from variation of the action (\ref{h14}) reads
\begin{equation}\label{h13}
(1+\lambda F)T^{\mu\nu}=G^{\mu\nu}-g^{\mu\nu}f+4H^{\mu\nu}(f^\prime+\lambda LF^\prime)
\end{equation}
where $F=F({\mathcal G})$ and $F^\prime=\frac{dF({\mathcal G})}{d{\mathcal G}}$. Taking the divergence of both side of this relation 
we obtain
\begin{eqnarray}\label{h51}
\nabla_\nu T^{\mu\nu}=\frac{\lambda F^\prime}{1+\lambda F}((g^{\mu\nu}L-T^{\mu\nu})\nabla_\nu{\mathcal G}+K^{\mu\nu}\nabla_\nu L)
\end{eqnarray}
where $K^{\mu\nu}=8RR^{\mu\nu}+4{R^\mu}_{\alpha\beta\gamma}R^{\nu\alpha\beta\gamma}-16R^{\mu}_\alpha 
R^{\alpha\nu}+8R^{\mu\alpha\beta\nu}R_{\alpha\beta}.$
The above equation can be used to obtain the matter equation of motion. For the perfect fluid described by energy-momentum tensor given in 
(\ref{h5}) projection parallel and normal to $u^\mu$ results in
\begin{eqnarray}\label{j7}
{\dot\rho}+(p+\rho)\theta&=&\frac{\lambda F^\prime}{1+\lambda F}\left((L+\rho)
{\dot\mathcal G}+K^{\mu\nu}u_\mu\nabla_\nu L\right)
\end{eqnarray}
and
\begin{eqnarray}\label{j1}
(p+\rho){\dot u^\mu}+\nabla^\mu p+u^\mu{ \dot p}&=&\frac{\lambda F^\prime h^\mu_\nu}{1+\lambda F}((L-p)\nabla^\nu {\mathcal G}\nonumber\\&&+K^{\nu\alpha}\nabla_\alpha L)
\end{eqnarray}
respectively. In the absence of non-minimal coupling the right hand side of this equation vanishes and we arrive at the geodesic equation.
The right hand side of equation (\ref{j1}) is non-vanishing in general which means that test particles feel an extra force which depends on 
$F({\mathcal G})$. This extra force which originates from the non-minimal coupling  is normal to the trajectories of test particles and results 
in non-geodesic motions. 

For a perfect dust with $L=-\rho$, equation (\ref{j1}) reduces to
\begin{equation}\label{j2}
{\dot u^\mu}=\frac{-\lambda F^\prime h^\mu_\nu}{1+\lambda F}(\nabla^\nu{\mathcal G}+K^{\nu\alpha}\nabla_\alpha \ln\rho).
\end{equation}
For the low energy effective action with matter part given by 
\begin{equation}\label{k1}
L=p(\phi,X)
\end{equation}
in which $\phi$ is a scalar field and $$X=\frac{1}{2}\nabla_\mu\phi\nabla^\mu\phi$$ the energy momentum tensor may be transformed into
the perfect fluid form given by (\ref{h5}) by setting $$u^\mu=\frac{\nabla^\mu\phi}{\sqrt{2X}},$$ provided $\nabla^\mu\phi$ being time-like
\cite{vik}. The equation of motion can then be obtained from (\ref{j1}).

Equation (\ref{h51}) may be compared with the one obtained in \cite{bertolami} for $f(R)$ gravity
\begin{eqnarray}\label{h53}
\nabla_\nu T^{\mu\nu}=\frac{\lambda f^\prime_2}{1+\lambda f_2}(g^{\mu\nu}L-T^{\mu\nu})\nabla_\nu R.
\end{eqnarray}
In the latter the perfect fluid Lagrangian $L=p$ results in a vanishing extra force but equation (\ref{j1}) shows that this is not the case for 
$f({\mathcal G})$ gravity. Similarly by choosing $L=-\rho$, the right hand side of (\ref{j7}) does not vanish but it vanishes for the 
counterpart equation in $f(R)$ gravity.
\section{World-line deviations}
In general relativity the equation of geodesic deviation is used to describe relative motions of test particles in a frame independent way.
Here we obtain its counterpart in $f({\mathcal G})$ gravity with non-minimal coupling to matter which we call the equation of world-line 
deviation. To this end, we consider a congruence of world-lines parametrized by $\epsilon$ and compare the equations of motion (\ref{j2}) for 
two particles moving on nearby trajectories $x^\mu(\tau)$ and $x^\mu(\tau)+\epsilon n^\mu(\tau)$ respectively, where $\epsilon$ is small and 
$n^\mu(\tau)=\frac{dx^\mu(\tau)}{d\epsilon}$. On keeping only terms which are linear in $\epsilon$ at most and defining $${\mathcal 
F}^\nu=\frac{-\lambda F^\prime}{1+\lambda F}(\nabla^\nu{\mathcal G}+K^{\nu\alpha}\nabla_\alpha \ln\rho),$$ we obtain
\begin{eqnarray}\label{h9}
{\ddot n}^\mu=-{R^\mu}_{\alpha\nu\beta}u^\alpha n^\nu u^\beta+h^\mu_\nu n^\alpha\nabla_\alpha{\mathcal F}^\nu+{\mathcal 
F}_\nu(u^\nu{\dot n}^\mu+u^\mu{\dot n}^\nu)
\end{eqnarray}
where we have used 
\begin{eqnarray}
\frac{Du^\mu}{D\epsilon}&=&\frac{Dn^\mu}{D\tau},\label{g4}\\
\frac{D}{D\epsilon}\frac{D}{D\tau}u^\mu&=&\frac{D}{D\tau}\frac{D}{D\epsilon}u^\mu+{R^\mu}_{\alpha\nu\beta}u^\alpha n^\nu u^\beta.\label{g5}
\end{eqnarray}
The first terms in the right hand side of equation (\ref{h9}) shows the usual effect of the background curvature as in the well-known geodesic 
deviation equation. The second term shows a contribution from the non-minimal coupling which is normal to the four-velocity and depends on 
$n^\mu$. The last term is proportional to the relative velocity. Such velocity-dependent terms also appear in the Palatini formulation of the 
$f(R)$ extended gravity without coupling to matter \cite{shoj}. The spacial components of (\ref{h9}) may written in the form of
\begin{equation}\label{f4}
{\vec a}={\vec a}_l+{\vec a}_t
\end{equation}
where $l,t$ stand for longitudinal and transverse with respect to the velocity. The transverse component is responsible for the deviation from 
the geodesic path. 

By following the procedure described above, one can obtain a similar deviation equation for $f(R)$ gravity with non-minimal coupling. Starting
from equation (\ref{h53}) we can obtain the following equation of motion for perfect dust particles
\begin{equation}\label{j21}
{\dot u^\mu}=\frac{-\lambda f^\prime_2 }{1+\lambda f_2}h^\mu_\nu\nabla^\nu R
\end{equation}
from which we arrive at equation (\ref{h9}) again, but with ${\mathcal F}^\nu$ replaced by $${\mathcal F}^\nu_2=\frac{-\lambda f^\prime_2 }
{1+\lambda f_2}\nabla^\nu R.$$ An interesting difference between ${\mathcal F}^\nu_2$ and ${\mathcal F}^\nu$ is that aside from a coupling 
constant $\lambda$ the former depends only on purely geometric objects but the latter depends on both geometric objects and matter mass density
directly. 

By an expansion of the form $F({\mathcal G})=F(0)+F^\prime(0){\mathcal G}+\cdots$, we obtain 
$${\mathcal F}^\nu=-\lambda F^\prime(0)(\nabla^\nu{\mathcal G}+K^{\nu\alpha}\nabla_\alpha\ln\rho)+\cdots$$  which is quadratic in curvature.
A similar expansion for the case of $f(R)$ gravity with non-minimal coupling results in an expression which is linear in curvature. From this 
vantage point, the deviation from geodesics is weaker in the model considered here. For large curvatures the deviation would be stronger in non-
minimal $f({\mathcal G})$ gravity.    

Equation (\ref{h9}) may be used to study the stability of orbits of particles in the framework of the model under consideration. Given a 
specific orbit of a particle in a space-time which is a solution of the equation of motion (\ref{h13}) (or its $f(R)$ gravity counterpart), one 
can solve equation (\ref{h9}) for the deviation $n^\mu$ from the fiducial orbit. This can then be put into  
\begin{equation}
\Lambda=\lim_{\tau\rightarrow\infty}\frac{1}{\tau}\ln\left(\frac{\sqrt{n_\mu(\tau)n^\mu(\tau)}}{\sqrt{n_\mu(0)n^\mu(0)}}\right)
\end{equation} 
to compute the relevant Lyapunov index which in turn is used to determine the stability behaviour of the orbits (see e.g. \cite{zo} for a 
discussion in the context of general relativity). Non-minimal coupling contributes to this through ${\mathcal F}^\mu$-dependent terms.
\section{Discussion}
In this work we showed that in $f({\mathcal G})$ gravity model with non-minimal coupling of an arbitrary function of the Gauss-Bonnet curvature 
to the matter Lagrangian, test particles move along non-geodesic trajectories due to the presence of extra forces originated from the coupling. 
These extra forces are perpendicular to the particles velocities and depend on both the choice of the function couples to matter and the matter 
Lagrangian. These results qualitatively agrees with those obtained in \cite{bertolami} for the case of non-minimal coupling in $f(R)$ gravity.
However, as we have shown, there are important distinctions between these models, namely the extra force in non-minimal $f(R)$ model vanishes 
for certain usual choices of matter Lagrangian but is does not in non-minimal $f({\mathcal G})$ model. We also generalized the geodesic 
deviation equation to a world-line deviation equation for both $f({\mathcal G})$ and $f(R)$ gravities with non-minimal couplings. According to 
this world-line deviation equation, relative accelerations of test particles depends on background curvature (as in general relativity), 
relative distances, and relative velocities of the particles. This is similar to the results obtained in the case of $f(R)$ gravity in its 
Palatini formulation without non-minimal coupling even though their origins are different. We showed that compared with non-minimal $f(R)$ 
gravity the deviations from geodesics is weaker in non-minimal $f({\mathcal G})$ gravity for small curvatures. It was also shown 
schematically how the world-line deviation equation we have obtained might be deployed to study the stability of orbits of particles in the 
extended gravity model considered here. More elaborate study of these relative motions would be possible by taking a generalized Raychaudhuri 
equation into account. The issue of energy conditions and stability would also be of interest in the context of the model considered here.  

\section*{Acknowledgements} The author would like to thank an anonymous referee of Physics Letters B for valuable comments.

\end{document}